\documentclass[osajnl,twocolumn,showpacs,10pt]{revtex4-1} 
\usepackage{amsmath,amssymb,graphicx}
\usepackage[english]{babel}

\usepackage{amsfonts} 
\usepackage{amsmath} 
\usepackage{braket} 

\usepackage{textcomp} 

\begin{document}

\title{Tuning Curve of Type-0 Spontaneous Parametric Down-Conversion}

\author{Stefan Lerch}
\email[Corresponding author: ]{stefan.lerch@iap.unibe.ch}
\author{B\"{a}nz Bessire}
\author{Christof Bernhard}
\author{Thomas Feurer}
\author{Andr\'{e} Stefanov}
\affiliation{University of Bern, Institute of Applied Physics, Sidlerstr. 5, 3012 Bern, Switzerland}

\begin{abstract}
We study the tuning curve of entangled photons generated by type-0 spontaneous
parametric down-conversion in a periodically poled KTP crystal. We demonstrate the X-shaped spatiotemporal structure of the spectrum by means of measurements and numerical simulations. Experiments for different pump waists, crystal temperatures, and crystal lengths are in good agreement with numerical simulations.
\end{abstract}

\maketitle 

\section{Introduction}
\label{sec:introduction}
Entangled photons are a primary source for studying the properties of entanglement \cite{Horodecki2009a}. They allow to perform fundamental tests of quantum mechanics \cite{Zeilinger1999,Genovese2005} and to implement protocols which could not be realized within the framework of classical physics. These cover quantum cryptography \cite{gisin2002} and more generally, quantum communication \cite{Gisin2007} and quantum computing \cite{Kok2007}. Applications like quantum imaging \cite{lugiato2002} or quantum optical coherence tomography \cite{nasr2003} revealed the potential of entangled photons to exceed classical resolution limits. The most common sources for entangled two-photon states are based on spontaneous parametric down-conversion (SPDC) occurring in nonlinear crystals \cite{klyshko1988}. This process allows to create entanglement with respect to polarization \cite{kwait1999}, angular orbital momentum \cite{mair2001}, momentum \cite{law2004} or energy \cite{rubin1994}.

Energy entanglement appears due to strong quantum correlations between the frequencies of the individual photons in a pair. Some applications require the spectrum of the individual photons to be broad \cite{nasr2008}. For instance, a broad spectrum allows to achieve a high axial resolution in quantum optical coherence tomography \cite{abouraddy2002} and the temporal properties of entangled photons can be measured with a spatial light modulator \cite{peer2005,zaeh2008}. Furthermore, the maximal flux of down-converted photons in the single photon regime is limited by the bandwidth of the entangled photon spectrum \cite{dayan2005}. In general, the interplay between the energy and momentum degrees of freedom within a two-photon state shows a complex spatiotemporal structure of entanglement \cite{Caspani2010, gatti2009}. Therefore, a detailed understanding of the SPDC process, as well as an accurate characterization of the down-converted spectrum is important. The joint spectrum of photon pairs has been measured for \mbox{type-I} \cite{Wasilewski2006,Baek2008a} and for \mbox{type-II} SPDC \cite{Kim2005,hendrych2007}. The \mbox{type-I} SPDC spectrum as a function of the pump waist is investigated in \cite{carrasco2006} and its transverse momentum dependency has been measured in the high gain regime 
\cite{lantz1993,devaux2000,Jedrkiewicz2011}. Coherence properties of the down-converted beam are investigated in \cite{hamar2010, jedrkiewicz2006}.

In this letter we study both theoretically and experimentally the full dependency of the type-0 SPDC spectrum on the transverse momentum \cite{Joobeur1994} for a monochromatic pump focused into a periodically poled crystal in the single photon regime. In the first part we theoretically describe the spectrum generated by SPDC and we identify in particular four parameters which determine the spectrum. We present an experimental setup which allows to measure the tuning curves, i.e.~the transverse momentum and frequency dependency of the SPDC emission, with respect to three of these parameters. Finally, we compare experimental results with numerical simulations.

\section{Theory}
\label{sec:theory}

Quasi-phase-matched (QPM) parametric amplification has been computed in \cite{bencheikh1995}. In the low gain regime the entangled two-photon state can equivalently calculated by first order perturbation theory as in \cite{Svozilik2009} where the generation of entangled photon pairs in periodically poled crystals has been theoretically studied. We consider SPDC induced by an undepleted monochromatic pump beam of angular frequency $\omega_{p}$ with a transverse field distribution $\mathcal{E}_p^+(\mathbf{q}_p)$, which propagates along the \mbox{z-axis} of a crystal with transverse momentum $\mathbf{q}_p$ of the wave vector $\mathbf{k}_p=(\mathbf{q}_p,k_{p,z})$. The pump photon $(p)$ is down-converted into the idler $(i)$ and signal $(s)$ photon with frequency \mbox{$\omega_i=\omega_{p}-\omega_s$} and $\omega_s$, respectively. All involved photons are identically polarized, i.e.~in type-0 configuration. Adapting the derivation of the entangled two-photon state in \cite{shih2003} for QPM type-0 configuration with finite crystal length yields to

\begin{widetext}
\begin{equation}
\ket{\Psi} = \ket{0}+\int\mathrm{d}^2q_i \int \mathrm{d}^2q_s \int \mathrm{d}\omega_s \; \Lambda(\mathbf{q}_i,\omega_{p}-\omega_s,\mathbf{q}_s,\omega_s) \hat{a}^\dagger_i(\mathbf{q}_i,\omega_{p}-\omega_s) \hat{a}^\dagger_s(\mathbf{q}_s,\omega_s)\ket{0},
\label{eq:psi}
\end{equation}
\begin{equation}
\Lambda(\mathbf{q}_i,\omega_{p}-\omega_s,\mathbf{q}_s,\omega_s) = -\dfrac{2i \epsilon_0\chi_0^{(2)} L(T) e(\omega_{p}-\omega_s) e(\omega_s)}{3\hbar(2\pi)^5 n(\omega_{p}-\omega_s,T) n(\omega_s,T)}\mathcal{E}_p^+(\mathbf{q}_i+\mathbf{q}_s)
\mathrm{sinc} \left[ \dfrac{\left( \Delta k_z+\frac{2\pi}{G(T)} \right) L(T)}{2} \right],
\label{eq:lambda}
\end{equation}
\end{widetext}

where $\ket{0}$ abbreviates the combined vacuum state $\ket{0}\doteq\ket{0_i}\ket{0_s}$. The photon creation operator is denoted by $\hat{a}^\dagger_j(\mathbf{q}_j,\omega_j)$ and $\Lambda(\mathbf{q}_i,\omega_{p}-\omega_s,\mathbf{q}_s,\omega_s)$ is the temperature dependent spectral amplitude function. Apart from the vacuum permittivity $\epsilon_0$ and the reduced Planck constant $\hbar$, the strength of the SPDC process is governed by the second order susceptibility $\chi_0^{(2)}$ and the length $L(T)$ of the nonlinear crystal. Dispersion properties are included through a frequency and temperature dependent refractive index $n(\omega_j,T)$ where we neglect its transverse wave vector dependencies. The phase mismatch

\begin{eqnarray}
\Delta k_z & = & \sqrt{\left(\dfrac{\omega_{p}-\omega_s}{c} n(\omega_{p}-\omega_s,T)\right)^2 -\mathbf{q}_i^2}\nonumber\\
&&+ \sqrt{\left(\dfrac{\omega_s}{c}n(\omega_s,T)\right)^2-\mathbf{q}_s^2} \nonumber\\
& & -\sqrt{\left(\dfrac{\omega_{p}}{c}n(\omega_{p},T)\right)^2-(\mathbf{q}_i+\mathbf{q}_s)^2},
\label{eq:dk}
\end{eqnarray}

is compensated by a judicious choice of the poling period $G(T)$ such that

\begin{equation}
\Delta k_z+\dfrac{2\pi}{G(T)}\cong 0.
\label{eq:phasematching}
\end{equation}

The normalization function

\begin{equation}
e(\omega_j)=i\sqrt{\dfrac{\hbar\omega_j}{2(2\pi)^3\epsilon_0c}},
\label{eq:fieldnormalization}
\end{equation}

originates from the field operator

\begin{eqnarray}
\hat{E}_j(\mathbf{q}_j,\omega_j,z) &=& \hat{E}_j^+(\mathbf{q}_j,\omega_j,z) +\hat{E}_j^-(\mathbf{q}_j,\omega_j,z)\nonumber\\
& =& e(\omega_j) \hat{a}_j(\mathbf{q}_j,\omega_j) \mathrm{e}^{ik_{j,z}z} + \mathrm{h.c.},
\label{eq:fieldoperator}
\end{eqnarray}

where $c$ is the speed of light in vacuum. The symmetry in \mbox{$\Lambda(\mathbf{q}_i,\omega_{p}-\omega_s,\mathbf{q}_s,\omega_s)$} with respect to \mbox{$\omega_{p}-\omega_s\leftrightarrow \omega_s$} and \mbox{$\mathbf{q}_i\leftrightarrow\mathbf{q}_s$} causes the indistinguishability of signal and idler spectra. Therefore, the total spectrum measured by a spectrometer can be considered to be proportional to the signal spectrum. Glauber's first order correlation function for the signal photon without the normalization function in the field operators \cite{glauber1963}

\begin{eqnarray}
S(\mathbf{q}_s,\omega_s,z) &=& G^{(1)}_{s}(\mathbf{q}_s,\omega_s,z)\nonumber\\
&=&\mathrm{Tr}\{\hat{\rho}_s \hat{E}_{s,z}^-(\mathbf{q}_s,\omega_s,z) \hat{E}_{s,z}^+(\mathbf{q}_s,\omega_s,z)\}
\label{eq:G1}
\end{eqnarray}

is proportional to the probability of measuring a photon with $\mathbf{q}_s$ and $\omega_s$ on a plane at the position $z$, i.e.~a momentum dependent spectral photon count density, henceforth referred to as spectral density at $z$. The operator \mbox{$\hat{\rho}_s = \mathrm{Tr_i}\{ \hat{\rho}\}$} is the partial trace over the idler subsystem of the density operator \mbox{$\hat{\rho}=\ket{\Psi}\bra{\Psi}$}. The assumption of a monochromatic pump beam and the field operator in Eq.~\eqref{eq:fieldoperator} without normalization function yield a $z$-independent spectral density

\begin{equation}
S(\mathbf{q}_s,\omega_s) = \int \mathrm{d}^2q_i \left| \Lambda(\mathbf{q}_i,\omega_{p}-\omega_s,\mathbf{q}_s,\omega_s) \right|^2.
\label{eq:spectrum}
\end{equation}

For a Gaussian beam given by

\begin{equation}
\mathcal{E}_p^+(\mathbf{q}_i+\mathbf{q}_s) \propto \exp\left(-\dfrac{w_0^2(\mathbf{q}_i+\mathbf{q}_s)^2}{4}\right),
\label{eq:gauss}
\end{equation}

the spectral density depends only on the absolute value of $\mathbf{q}_s$. The poling period $G$, pump beam waist $w_0$, temperature $T$ and crystal length $L$ are the relevant parameters for the shape of $S(\mathbf{q}_s,\omega_s)$. As reported in \cite{feyer1992} the temperature affects not only the refractive index but also $G(T)$ and $L(T)$ by thermal expansion $G(T)=G_0\left[1+\alpha(T-25^\circ\mathrm{C})+\beta(T-25^\circ\mathrm{C})^2\right]$ with coefficients $\alpha$ and $\beta$ taken from \cite{emanueli2003}. This equation applies also for $L(T)$. All numerically simulated tuning curves are
done by calculating Eq.~\eqref{eq:spectrum} for a given variable set $(\mathbf{q}_s, \omega_s)$. The integration boundaries
for $\mathbf{q}_i$ are limited to an empirically found range where the phase matching condition is
fulfilled.

\section{Experimental Setup}
\label{sec:experiment}

\begin{figure}[htbp]
\centering
\includegraphics[width=1\linewidth]{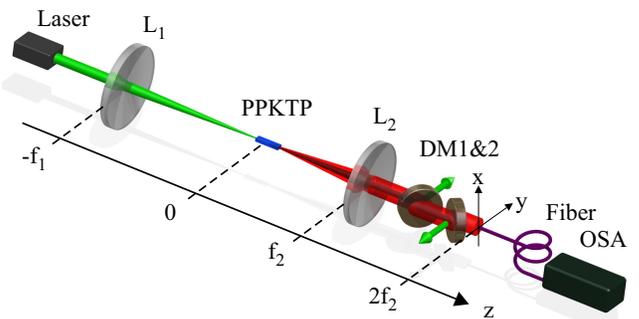}
\caption{(Color online) Schematics of the experimental setup for measuring the spectral density.}
\label{fig:setup}
\end{figure}

\begin{figure}
\centering
\includegraphics[width=1\linewidth]{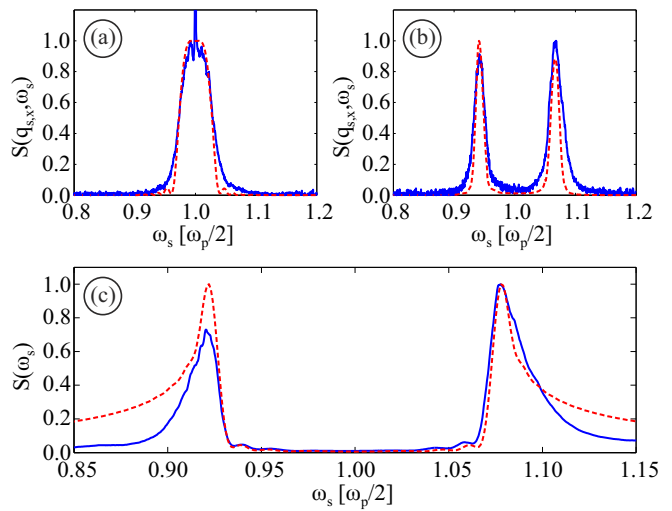}
\caption{(Color online) Typical normalized measured (solid curve) and simulated (dashed curve) spectral densities for two different fiber positions (\mbox{x = 0 mm} (a), \mbox{x = 1 mm} (b)). The narrow peak in (a) at $\omega_{p}/2$ comes from remaining light of the Verdi which is neither frequency doubled nor filtered. In (c) the fiber at \mbox{x = 0 mm} is replaced by a collimator. The measured (solid curve) spectral density shows the same structure at $0.95\times\omega_{p}/2$ and $1.05\times\omega_{p}/2$ as the simulation (dashed curve), originating from the sinc term in Eq.~\eqref{eq:lambda}. The asymmetry in the measured curve arises because a photon with frequency $\omega_p/2-\omega_0$ (\mbox{$\omega_0>0$ rad/s}) diverges more than its twin photon with frequency $\omega_p/2+\omega_0$, giving rise to a lower coupling efficiency.}  
\label{fig:sinc}
\end{figure}

The setup used to measure the spectral density of Eq.~\eqref{eq:spectrum} is depicted in Fig.~\ref{fig:setup}. The pump laser is a quasi-monochromatic Nd:YVO$_4$ (Verdi) with central wavelength $\lambda_{p}=532$~nm and a spectral bandwidth of \mbox{$5$~MHz}. A pump power of 5~W is focused by a lens $L_1$ with either \mbox{$f_1=300$~mm} or \mbox{$f_1=150$~mm} into a periodically poled potassium titanyl phosphate (PPKTP) crystal where SPDC generates entangled photons centered around $\lambda_{c}=1064$~nm. The down-converted photon power is measured to be linear to the pump power and around $P_{DC}=400$~nW (parameter dependent) which correspond to a flux of \mbox{$\Phi_{DC}=2.14$ $\times$ $10^{12}$ photons/s}. According to \cite{dayan2005} the maximal flux of down-converted photons $\Phi_{max}$ that can still be considered as composed of distinct photon pairs is approximately the down-converted bandwidth $\Delta_{DC}$ which is in our case greater than \mbox{50 nm}. This leads to an experimental spectral mode density $n_{DC}=\Phi_{DC}/\Phi_{max}<0.16$ which confirms the single photon limit and legitimate the quantum mechanical description of our experiment. The parameter $w_0$ can be varied by changing the focal length $f_1$. The SPDC source is mounted in a copper block whose temperature is controlled to $\pm 0.1$~$^\circ$C. Heat transfer between copper and PPKTP is ensured by wrapping the crystal with an indium foil. Due to the low absorption coefficient of KTP at $\lambda_p$, the temperature gradient between the location of the absorbed pump power within the crystal and the crystal surface is negligible as simulations have shown. Therefore, the measured parameter $T$ of the copper block can be considered to be equal to the effective temperature of the crystal. Two different crystals are used to analyze the effect on Eq.~\eqref{eq:spectrum} of different lengths $L$. The pump beam is filtered out by two dichroic mirrors (DM1\&2) oriented such that the beam displacement is compensated. The wave vector-dependent spectrum is measured by raster-scanning a multi-mode fiber along the x-axis of a 2f imaging system with a lens $L_2$ of focal length \mbox{$f_2=50mm$}. The fiber position $x$ is related to the transverse momentum by $q_{s,x}=\omega_s x/(f_2c)$. Since $S(\mathbf{q}_s,\omega_s)=S(|\mathbf{q}_s|,\omega_s)$ it is sufficient to scan the fiber along the x-axis. The direction of the remaining light from the pump beam after the two dichroic mirrors defines the optical axis and is used to align the fiber. The fiber is subsequently coupled into an optical spectrum analyzer (OSA) with \mbox{70~dB} dynamic range. Two typical spectral densities are indicated in Fig.~\ref{fig:sinc} (a) and (b) for different fiber positions. The measured spectra are systematically broader than the simulated ones which is caused by limited resolution. Limiting factors are finite fiber core of \mbox{$200$~\textmu m}, chosen OSA resolution of \mbox{2~nm} and imperfections in the 2f-image such as refraction at crystal surface, astigmatism of $L_2$, and uncertainty about the distance between $L_2$ and fiber. The finite resolution smooths out the sinc structure given by Eq.~\eqref{eq:lambda}. However for mostly collinear but non-degenerated cases the sinc structure is visible when measuring the total spectral density by replacing the fiber by a collimator as shown in Fig.~\ref{fig:sinc} (c). Only in the non-degenerated case small side lobe contributions of the sinc structure remain after integration over $\mathbf{q}_s$ as can be seen in Fig.~\ref{fig:numericalsimulation}. 

\section{Results and Discussion}
\label{sec:results and discussion}
In the following the influence of the four parameters $G$, $w_0$, $T$, and $L$ is investigated independently.

\subsection{Poling Period $G$}

\begin{figure}[htbp]
\centering
\includegraphics[width=1\linewidth]{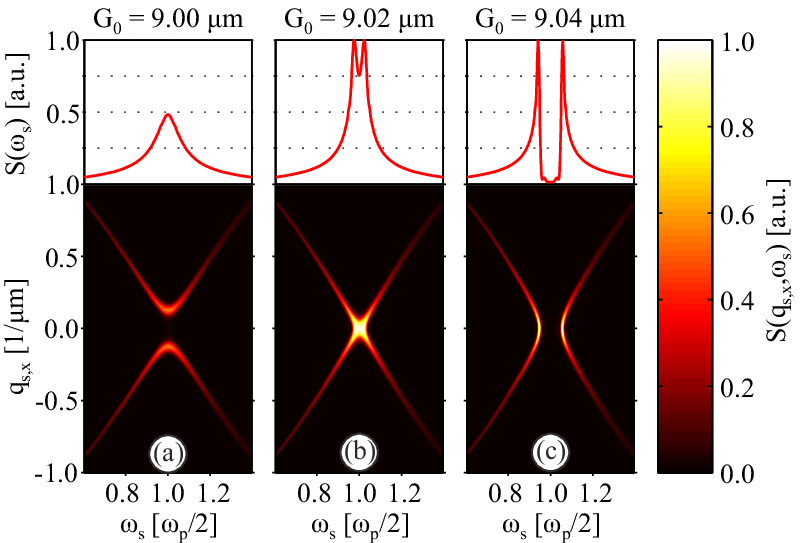}
\caption{(Color online) Simulation of the spectral density Eq.~\ref{eq:spectrum} (normalized to the maximum of (b)) generated by a Gaussian beam in a PPKTP crystal. The parameters are \mbox{$w_0=23.27$ \textmu m}, \mbox{$T=25$ ${}^\circ$C}, \mbox{$L_0=7.5$ mm}, and (a) \mbox{$G_0=9.00$ \textmu m}, (b) \mbox{$G_0=9.02$ \textmu m}, and (c) \mbox{$G_0=9.04$ \textmu m}. The plots on the top result from integration along the \mbox{$q_{s,x}$-axis}.}
\label{fig:numericalsimulation}
\end{figure}

The poling period $G_0$ is not accessible experimentally and, therefore, we consider its influence on the spectral density in Eq.~\eqref{eq:spectrum} only by numerical simulations with parameters taken from the experiment (see Fig.~\ref{fig:numericalsimulation}). The tuning curve $S(|\mathbf{q}_s|,\omega_s)$ shows an overall X-shaped structure with a maximum around $\omega_{p}/2$ and small $|\mathbf{q}_s|$ values. The symmetry in the width of the branches with respect to a line at $\omega_{p}/2$ is not exact because the refractive index is frequency dependent. From Eq.~\eqref{eq:phasematching} follows that spectral components propagate in different directions, giving rise to the X-shaped structure and thus colored emission rings. The emission angle relative to the optical axis can be calculated from $\omega_{s}$ and $|\mathbf{q}_s|$ by $\theta=\arcsin\left(|\mathbf{q}_s| / |\mathbf{k}_s|\right)$, which implies a larger emission angle for the photon with lower frequency.

A change of $G_0$ affects the phase matching condition in Eq.~\eqref{eq:phasematching} which leads to another propagation direction for the different spectral components. For \mbox{$G_0=9.00$ \textmu m} very few photons propagate parallel to the optical axis, but most of the photons have frequency $\omega_p/2$, corresponding to a non-collinear but degenerate emission. With a slightly longer poling period (\mbox{$G_0=9.02$ \textmu m}) the emission probability favors the collinear case and the overall down-conversion efficiency increases. For even longer poling periods (\mbox{$G_0=9.04$ \textmu m}) the emission still favors the collinear case, however, becomes \mbox{non-degenerate}; thus a gap at $\omega_{p}/2$ in the total spectrum can be observed. Further enhancement of $G_0$ broadens the gap and leads to additional local maxima inside the gap due to the sinc in Eq.~\eqref{eq:spectrum}. The simulations show that variations of $G_0$ in the nm-regime strongly affect the spectral density. Since $G_0$ is typically not specified to such precision, it is adjusted such that the simulations agree with the experimental results.

\subsection{Pump Beam Waist $w_0$}

\begin{figure}[htbp]
\centering
\includegraphics[width=1\linewidth]{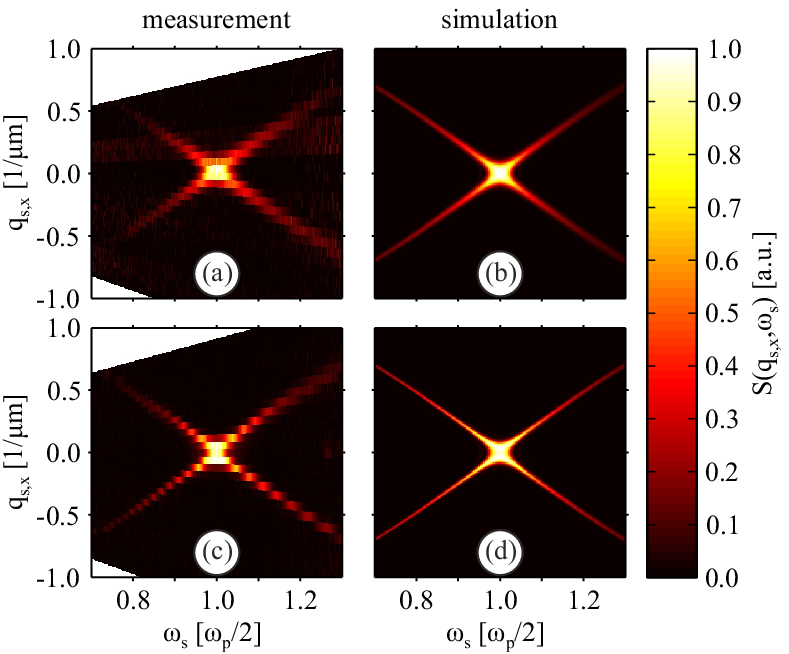}
\caption{(Color online) Measurement, (a) and (c), and simulation, (b) and (d), of the normalized spectral density for different beam waists $w_0$. (a) and (b) \mbox{$f_1=150$ mm} and \mbox{$w_0=23.27$ \textmu m}, (c) and (d) \mbox{$f_1=300$ mm} and \mbox{$w_0=46.53$ \textmu m}. The other parameters are \mbox{$L_0=7.5$ mm}, \mbox{$T=25$ ${}^\circ$C}, and \mbox{$G_0=9.018$ \textmu m}.}
\label{fig:beam waist}
\end{figure}

The simulations for different beam waists in Fig.~\ref{fig:beam waist} show again a X-structure. Comparing the width of the branches illustrates that, if the pump beam is focused more tightly, each frequency is generated over a broader $|q_{s,x}|$-range. For the utilized crystal we get best agreements between simulation and measurement for $G_0=9.018$ \textmu m. All numerical and experimental data in Fig.~\ref{fig:beam waist} are normalized to their maximum value since the fiber coupling efficiency varies for both measurements. In fact, the measured signal decreases with a bigger beam waist since $w_0$ appears in the exponent in Eq.~\eqref{eq:gauss}. If the beam waist is too small the SPDC efficiency would decrease again since the beam divergence increases. The slight asymmetry in the width of the branches with respect to a line at $\omega_{p}/2$ is visible in both simulation and measurement and is again due to the frequency dependent refractive index. Apart from the efficiency the beam waist does not change the spectral density significantly as long as transverse wave vector dependencies in $n(\omega,T)$ can be neglected.\\

\subsection{Crystal Temperature $T$}

\begin{figure}[htbp]
\centering
\includegraphics[width=1\linewidth]{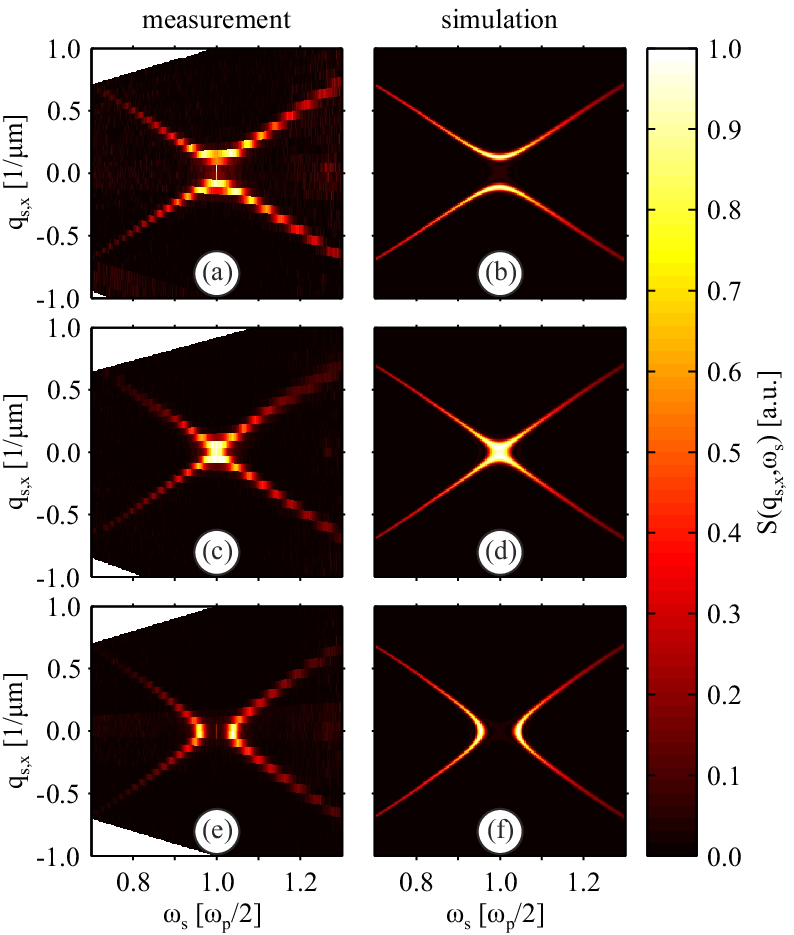}
\caption{(Color online) Measurements, (a), (c) and (e), and simulations, (b), (d) and (f), of the normalized spectral density for \mbox{$T=15$ ${}^\circ$C} ((a) and (b)), \mbox{$T=25$ ${}^\circ$C} ((c) and (d)), and \mbox{$T=35$ ${}^\circ$C} ((e) and (f)). The other parameters are \mbox{$w_0=46.53$ \textmu m}, \mbox{$L_0=7.5$ mm}, and \mbox{$G_0=9.018$ \textmu m}.}
\label{fig:temperature}
\end{figure}

By thermal expansion the temperature can change $G(T)$ and $L(T)$, and, in addition it influences the refractive index $n(\omega,T)$. Since $\partial n(\omega,T)/\partial T$ is small, the refractive index in the denominator of Eq.~\eqref{eq:lambda} has negligible influence on $S(|\mathbf{q}_s|,\omega_s)$. Conversely, $n(\omega,T)$ affects the phase mismatch in Eq.~\eqref{eq:dk} in a very sensitive manner. Figure~\ref{fig:temperature} illustrates measured and simulated results for various temperatures $T$. For \mbox{$T=15$ ${}^\circ$C} no spectral component propagates parallel to the optical axis, corresponding to a non-collinear and degenerate emission. By increasing the temperature the propagation direction of frequencies around $\omega_{p}/2$ begins to approach the pump beam direction. At \mbox{$T=25$ ${}^\circ$C}, the emission is mostly collinear and degenerate. Finally at \mbox{$T=35$ ${}^\circ$C}, the phase mismatch for $\omega_{p}/2$ is too high and the efficiency decreases; the emission is still collinear but non-degenerate. When comparing Fig.~\ref{fig:numericalsimulation} and Fig.~\ref{fig:temperature} it can be observed, as expected, that the phase matching can be tuned either by varying the temperature or by changing $G_0$. Increasing the temperature by \mbox{10 ${}^\circ$C} has the same effect as an elongation of the poling period of approximately \mbox{20 nm}. Due to thermal expansion, $G$ elongates about \mbox{0.6 nm} and therefore contributes only 3\% to changes in the phase matching condition (Eq.~\eqref{eq:phasematching}). The main contribution comes from the temperature depending refractive index entering in $\Delta k_z$. The variation of the spectrum due to a temperature dependent change of $L$ can be completely neglected, as numerical simulations confirmed. The measured spectral density for \mbox{$T=15$ ${}^\circ$C} is slightly asymmetric at $\omega_{p}/2$ with respect to a line at \mbox{$q_{s,x}=0$ \textmu m$^{-1}$}. The reason for this is the ring shaped spatial mode of the entangled photon beam in the non-collinear case which complicates the exact localization of the optical axis. Again experimental and numerical data are normalized to their maximal value. For simulations the temperature has to be known with a certainty of about 0.5 ${}^\circ$C.

\subsection{Crystal Length $L$}

\begin{figure}[htbp]
\centering
\includegraphics[width=1\linewidth]{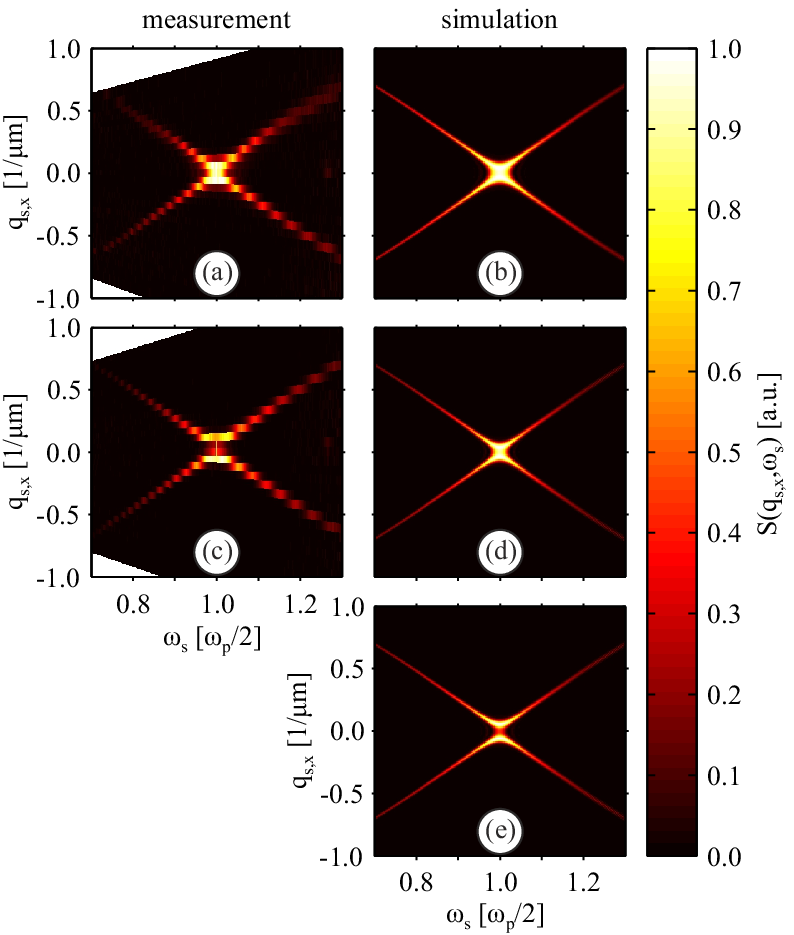}
\caption{(Color online) Measurements, (a) and (c), and simulations, (b), (d), and (e), of the normalized spectral density for \mbox{$L_0=7.5$ mm} ((a) and (b)) and \mbox{$L_0=12$ mm} ((c), (d), and (e)). The other parameters are \mbox{$w_0=46.53$ \textmu m}, \mbox{$T=25$ ${}^\circ$C}, and \mbox{$G_0=9.018$ \textmu m}. In (e) the poling period is changed to \mbox{$9.016$ \textmu m}.}
\label{fig:length}
\end{figure}

The measurements for two crystals from the same manufacturer and with same nominal poling period $G_0$ but of different lengths indicate a clear difference in $S(q_{s,x},\omega_s)$ as shown in Fig.~\ref{fig:length}. In contrast, the simulations indicate that a change in length only could not cause the emission to change from collinear to non-collinear as measured, but has merely the effect of narrowing the spectrum. The discrepancy in the measurement is explained by slightly different $G_0$ for the two crystals. In order to achieve better agreement between measured and simulated spectrum, $G_0$ for the longer crystal has to be decreased by \mbox{$2$ nm} in the simulation. The high sensitivity of $S(|\mathbf{q}_s|,\omega_s)$ on $G_0$ preclude an independent experimental investigation of the influence of $L_0$ on the spectral density.

\section{Conclusion}
\label{sec:conclusion}

We have investigated the spectral photon count density of type-0 entangled photons in a PPKTP crystal depending on their transverse momentum. We further have characterized how different parameters influence the spectrum and we reported good agreement between measurement and simulated data. It turned out that the spectrum itself is most sensitive to the crystals poling period $G_0$. A change in the nm-range causes different emission. In practice, due to the equivalent behavior of the spectral density when changing the temperature $T$ or $G_0$, the discrepancy between the nominal and effective poling periods can be compensated by temperature tuning. Increasing the temperature by \mbox{10 ${}^\circ$C} has the same effect as an elongation of the poling period of approximately \mbox{20 nm}. Thus, for simulations $G_0$ has to be known with nm accuracy and $T$ up to \mbox{0.5 ${}^\circ$C}. Low temperatures cause non-collinear and degenerate emission. Increasing $T$ the emission gets mostly collinear. For even higher temperature the emission still favors the collinear case but becomes non-degenerate. Moreover, we have noticed that, apart from the total SPDC efficiency, changes of the pump waist or of the crystal length do not influence critically the emission spectra. Hence, the nominal value of the length and a beam waist estimation are precise enough for accurate simulations. By means of momentum dependent spectral measurements, we have shown a good agreement between computation and measurement. The understanding of the spatiotemporal structure of the spectrum is of importance for all applications where entanglement in energy is the relevant degree of freedom and allows to optimize the spectrum for specific applications.

\section*{Acknowledgments}
This work was supported by the National Centre of Competence in Research - Molecular Ultrafast Sciences and Technology, and Swiss National Science Foundation grant PP00P2\_133596.



\bibliographystyle{osajnl}


\end{document}